\documentclass[traditabstract]{aa}

\usepackage{amsmath}
\usepackage{graphicx}
\usepackage{xspace}
\usepackage{natbib}
\usepackage{txfonts}
\usepackage{hyperref}
\bibpunct{(}{)}{;}{a}{}{,}

\newcommand{\suz}{\textsl{Suzaku}\xspace}
\newcommand{\bat}{\textsl{Swift}/BAT\xspace}
\newcommand{\nustar}{\textsl{NuSTAR}\xspace}
\newcommand{\integral}{\textsl{INTEGRAL}\xspace}
\newcommand{\rxte}{\textsl{RXTE}\xspace}

\newcommand{\snr}{S/N\xspace}

\newcommand{\ks}{KS~1947+300\xspace}
\graphicspath{{./figures/}}
\begin{document}

\title{\suz observations of the 2013 outburst of \ks}

\author{
     Ralf Ballhausen\inst{1} 
\and Matthias K\"uhnel\inst{1} 
\and Katja Pottschmidt\inst{2,3} 
\and Felix F\"urst\inst{4} 
\and Paul B. Hemphill\inst{5}
\and Sebastian Falkner\inst{1}
\and \mbox{Amy M. Gottlieb\inst{2,3}}
\and Victoria Grinberg\inst{6}
\and Peter Kretschmar\inst{7}
\and Ingo Kreykenbohm\inst{1}
\and Richard E. Rothschild\inst{5}
\and \mbox{J\"orn Wilms}\inst{1}
}

\authorrunning{Ballhausen et al.}

\institute{
	Dr.\ Karl-Remeis-Sternwarte and Erlangen Centre for
        Astroparticle Physics, Sternwartstr.~7, 96049 Bamberg, Germany 
\and 
	Department of Physics, University of Maryland Baltimore County,
Baltimore, MD 21250, USA
\and
	CRESST and NASA Goddard Space Flight Center, Astrophysics Science
Division, Code 661, Greenbelt, MD 20771, USA
\and 
	Cahill Center for Astronomy and Astrophysics, California Institute of Technology, Pasadena, CA 91125, USA
\and 
        Center for Astrophysics and Space Sciences, University of California, San Diego, 9500 Gilman Dr., La Jolla, CA 920093-0424, USA
\and
        Kavli Institute for Astrophysics and Space Research,
        Massachusetts Institute of Technology, Cambridge, MA 02139,
        USA 
\and    
Science Operations Division, Science Operations Department of
ESA, ESAC, Villanueva de la Ca\~nada (Madrid), Spain 
}

\abstract{We report on the timing and spectral analysis of two \suz
  observations with different flux levels of the high-mass X-ray
  binary \ks during its 2013 outburst. In agreement with simultaneous
  \nustar observations, the continuum is well described by an absorbed
  power law with a cut-off and an additional black body component. In
  addition we find fluorescent emission from neutral, He-like, and
  even H-like iron. We determine a pulse period of $\sim$18.8\,s with
  the source showing a spin-up between the two observations. Both \suz
  observations show a very similar behavior of the pulse profile,
  which is strongly energy dependent, with an evolution from a profile
  with one peak at low energies to a profile with two peaks of
  different widths towards higher energies seen in both, the \suz and
  \nustar data. Such an evolution to a more complex profile at higher
  energies is rarely seen in X-ray pulsars, most cases show the
  opposite behavior. Pulse phase-resolved spectral analysis shows a
  variation in the absorbing column density, $N_\mathrm{H}$, over
  pulse phase. Spectra taken during the pulse profile minima are
  intrinsically softer compared to the pulse phase-averaged spectrum. }

\date{Received DATE  / Accepted DATE}

\maketitle

\section{Introduction}

\object{KS~1947+300} is an accretion powered X-ray pulsar in the
constellation Cygnus. It was discovered on 1989 June 8 by the TTM
instrument on the \emph{Kvant} module of the Mir space station at a
flux level of $70\pm10\,\mathrm{mCrab}$ in the 2--27\,keV band
\citep{Borozdin1990}. Its energy spectrum can be described by an
absorbed power law of photon index $1.72\pm0.31$ and an equivalent
hydrogen column density of $(3.4\pm 3.0)\times
10^{22}\,\mathrm{cm}^{-2}$.

In 1994, this source was rediscovered as GRO~J1948+32 by the Burst and
Transient Science Experiment on the Compton Gamma-Ray Observatory. It
was identified as an X-ray pulsar with a pulse period of
$\sim$18.7\,s. This observation also allowed for a first estimate of
the orbital parameters \citep{Chakrabarty1995}. Further \rxte
observations of an outburst in 2000--2001 proved that \ks and
GRO~J1948+32 are the same object \citep{Levine2000, Swank2000}. Its
orbit was found to have an eccentricity
$e=0.033\pm0.013$, an orbital period
$P_\mathrm{orb}=40.415\pm0.010$\,d, and a semi-major axis of $a\sin
i=137.4$\,lt-s \citep{Galloway2004}. The optical companion was
identified as a B0Ve star \citep{Negueruela2003}.

\citet{Naik2006} presented a broad band spectral analysis using three
\textsl{BeppoSAX} observations of the 2000 November--2001 June
outburst. Their spectral model consisted of a Comptonized continuum, a
black body component with a temperature of $\sim$0.6\,keV, and an iron
emission line at 6.7\,keV. \citet{Tsygankov2005} reported on \integral
observations taken from 2002 December to 2004 April and presented
studies on the evolution of the pulse profile and period. They found a
proportionality of the pulsation frequency and flux near the peak of
the outburst, although at low significance, and they also observed a flux
dependence of the pulse profile shape.

In this paper we report on the analysis of two \suz observations from
2013 October and 2013 November taken during the first outburst of this
source since 2004. These observations were quasi simultaneous to
\nustar observations in which \citet{Furst2014} present evidence for
the presence of a Cyclotron Resonant Scattering Feature (CRSF or
cyclotron line) at 12.2\,keV. The detection of CRSFs allows us to
directly infer the magnetic field strength in the emission region,
which is a fundamental parameter of the neutron star \citep[see,
  e.g.,][for a review]{Harding2006}. \suz provides high sensitivity at
low energies and therefore allows us to study the soft component of
the broad band continuum, which is not available from \nustar alone,
including the absorption and iron fluorescence emission.

The paper is organized as follows: Section~\ref{sec:observation} gives
an overview of the data acquisition and reduction. In
Sect.~\ref{sec:timing}, we investigate the shape, energy dependence,
and evolution of the pulse profiles. Phase-averaged and phase-resolved
spectral analyses are presented in Sect.~\ref{sec:phase_averaged} and
\ref{sec:phase_resolved}, respectively. In Sect.~\ref{sec:discussion},
we summarize and discuss our results.

\section{Observation and Data Reduction}\label{sec:observation}

\subsection{\suz}
\suz is a joint mission of JAXA and NASA. It carries two main
detectors: The X-ray Imaging Spectrometer
\citep[XIS;][]{Takahashi2007} and the Hard X-ray Detector
\citep[HXD;][]{Koyama2007}. XIS consists of four individual but
identical Type-I Wolter telescopes (XRT), focusing the X-rays on CCD
cameras. The XIS1 camera is back-illuminated, providing higher
sensitivity at lower energies than the other front-illuminated
configurations with a maximum energy range of 0.2--12\,keV for all XIS
\citep{Takahashi2007}. In 2006, XIS2 was damaged by a micro-meteorite
impact and became unusable.  HXD consists of a collimated PIN silicon
diode array (PIN) and a GSO/BGO phoswitch counter (GSO), covering a
total energy range of 10--600\,keV \citep{Koyama2007}.

\begin{figure}
   \resizebox{\hsize}{!}{\includegraphics{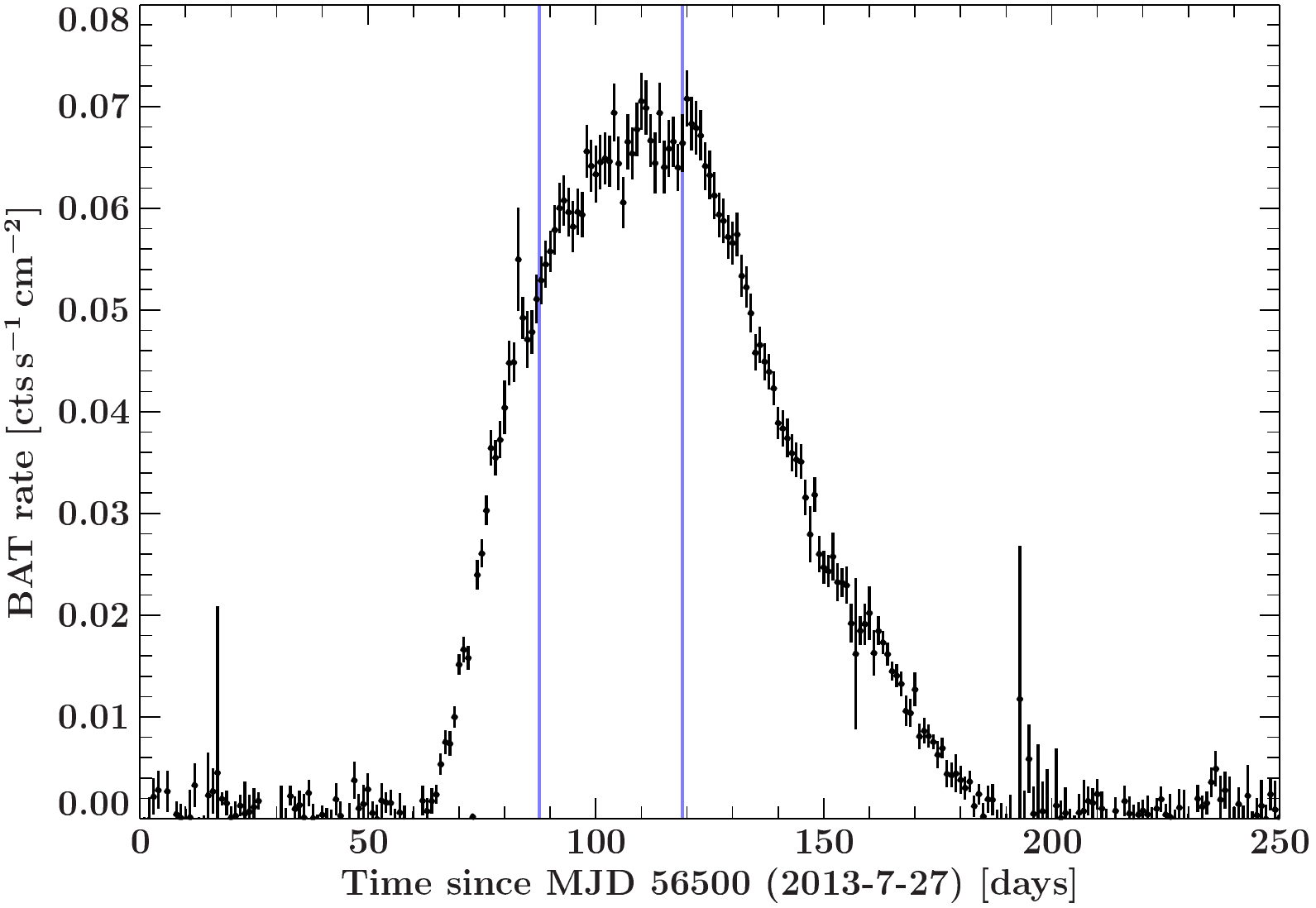}}
   \caption{\bat daily lightcurve in the 15--50\,keV band of the
     outburst in 2013. The blue vertical lines indicate the times
     covered by \suz observations.}
   \label{fig:bat_lightcurve}
\end{figure}

\suz observed the 2013 outburst of \ks twice, approximately one month
apart and partially coinciding with \nustar observations. The
observation log is summarized in Table~\ref{tab:obslog}.
Figure~\ref{fig:bat_lightcurve} shows the \bat lightcurve of \ks
\citep{Krimm2013} with times of the \suz observations marked. During
both observations, XIS was operated in $1/4$ Window mode and Clocking
modes were set to ``normal'' and ``burst'' for the first (Obs.~I) and
second observations (Obs.~II), respectively.

We used the software package HEAsoft (v.\ 6.15.1) for all \suz
reprocessing and extraction. Standard screening criteria and
calibration were applied by running \texttt{aepipeline} on both
datasets. All event times were transferred to the solar barycenter
using \texttt{aebarycen}. We used \suz CALDB v20110630 for XRT,
v20110913 for HXD, and v20130916 and v20131231 for the first and
second XIS data set, respectively.

Most likely due to incorrect attitude determination, the conversion of
XIS detector to sky coordinates yields incorrect values for the first
observation. As a result the sky image can not be properly
reconstructed and shows two distinct sources. We therefore performed
all data extractions for this observation in detector coordinates and
regions are given in units of pixels. This method does not allow an
additional attitude correction with the FTOOL \texttt{aeattcor2}
\citep[see][for details]{Uchiyama2008}. As the variation of the
effective area with off axis angle is small, the systematic error
introduced by this choice is small and does not affect our analysis.

For the second observation \texttt{aeattcor2} was used for further
improvement of the spacecraft's attitude. We used annular extraction
regions for XIS with outer radii of 125\,pixels (Obs.~I) and
$1\farcm5$ (Obs.~II). Inner radii were set individually to exclude all
regions with pile-up fractions larger than 4\% (65--72\,pixels for
Obs.~I and $35\arcsec$ for Obs.~II), which were estimated using
\texttt{pileest}. The XIS lightcurves were extracted with a 2\,s time
resolution, which is the highest possible resolution for the selected
operating modes.

We applied the PIN response for calibration epoch 11 for the
XIS nominal pointing position. We used the PIN ``tuned'' background
(v2.2) as the non X-ray background model \citep{Fukazawa2009} for all
analyses and also took the cosmic X-ray background into account
\citep{Boldt1987}. We applied the GSO ``correction ARF'', provided by
the \suz HXD team to improve the calibration of this instrument
\citep[see][for details]{Yamada2011} to both observations. PIN has a
nominal time resolution of $61\,\mu\mathrm{s}$. Because the good time
intervals corresponding to the phase bins are too short for an
individual correction, the overall loss of artificially triggered
``pseudo'' events was used to estimate the dead time for the pulse
profiles.

We extracted lightcurves and spectra for all available \suz
instruments using \texttt{xselect} for the XIS, and the \suz specific
FTOOLs \texttt{hxdpinxbpi}/\texttt{hxdpinxblc} and
\texttt{hxdgsoxbpi}/\texttt{hxdgsoxblc} for the extraction of the PIN
and the GSO data, respectively.

In order to avoid well known uncertainties in the effective area and
response matrix determination, we considered only the energy range of
1--10\,keV for XIS0 and XIS3 and 1--8\,keV for XIS1 for the spectral
analysis. The energy ranges of 1.72--1.88\,keV and 2.19--2.37\,keV
were excluded for all XIS due to known Au and Si calibration features
\citep{Nowak2011}. For PIN and GSO we used the energy ranges of
15--70\,keV and 70--90\,keV, respectively. We used the
\emph{Interactive Spectral Interpretation System} \citep[ISIS v1.6.2,
][]{Houck2000}. Uncertainties are given at the 90\% confidence level
(CL) for one parameter of interest unless otherwise noted.

\begin{table}
\caption{Observation log for the two \suz and the \nustar observation.
  Uncertainties of the pulse period are given at the 68\% confidence
  level.} 
\label{tab:obslog}
\centering
\begin{tabular}{ccccc}
\hline
Obs. No. & ObsID & mid-time & exposure & pulse period\\
 & & [MJD] & [ks] &  [s] \\
\hline\hline
\multicolumn{5}{c}{\suz} \\
\hline
Obs.~I & 908001010 & 56587.58 & 29.0 & 18.80876(7)\\
Obs.~II & 908001020 & 56619.02 & 7.6 & 18.78896(7)\\
\hline
\multicolumn{5}{c}{\nustar} \\
\hline
Obs.~III & 80002015004 & 56619.14  & 18.6 & --\\
\hline
\end{tabular}
\end{table}

\subsection{\nustar}
As mentioned above, simultaneous to our \suz observations there were
also two \nustar observations which have previously been published by
\citet{Furst2014}. We use these data to better illustrate the
interpretation of the pulse profile evolution with energy than what is
possible with \suz alone (Sect.~\ref{sec:timing}). We extracted the
\nustar data from ObsID 80002015004 using the standard pipeline
\texttt{nupipline} v.\ 1.4.1 as distributed with HEAsoft v.\ 6.16 and
CALDB v.\ 20150316. We used the same extraction regions as
\citet{Furst2014}, i.e., radii of $130\arcsec$ and $105\arcsec$, for
the source and background, respectively. Light curves were extracted
with 0.5\,s time-resolution. This time resolution is nominally faster
than the available dead time calculation, which is only available on a
1\,s basis. However, \ks is varying smoothly on these timescales so
that averaging over 1\,s bins for the dead time does not introduce
significant errors in the count rate estimate. The extracted
lightcurves were barycentered to the solar system.

\section{Timing analysis}\label{sec:timing}

\begin{figure}
 \resizebox{\hsize}{!}{\includegraphics{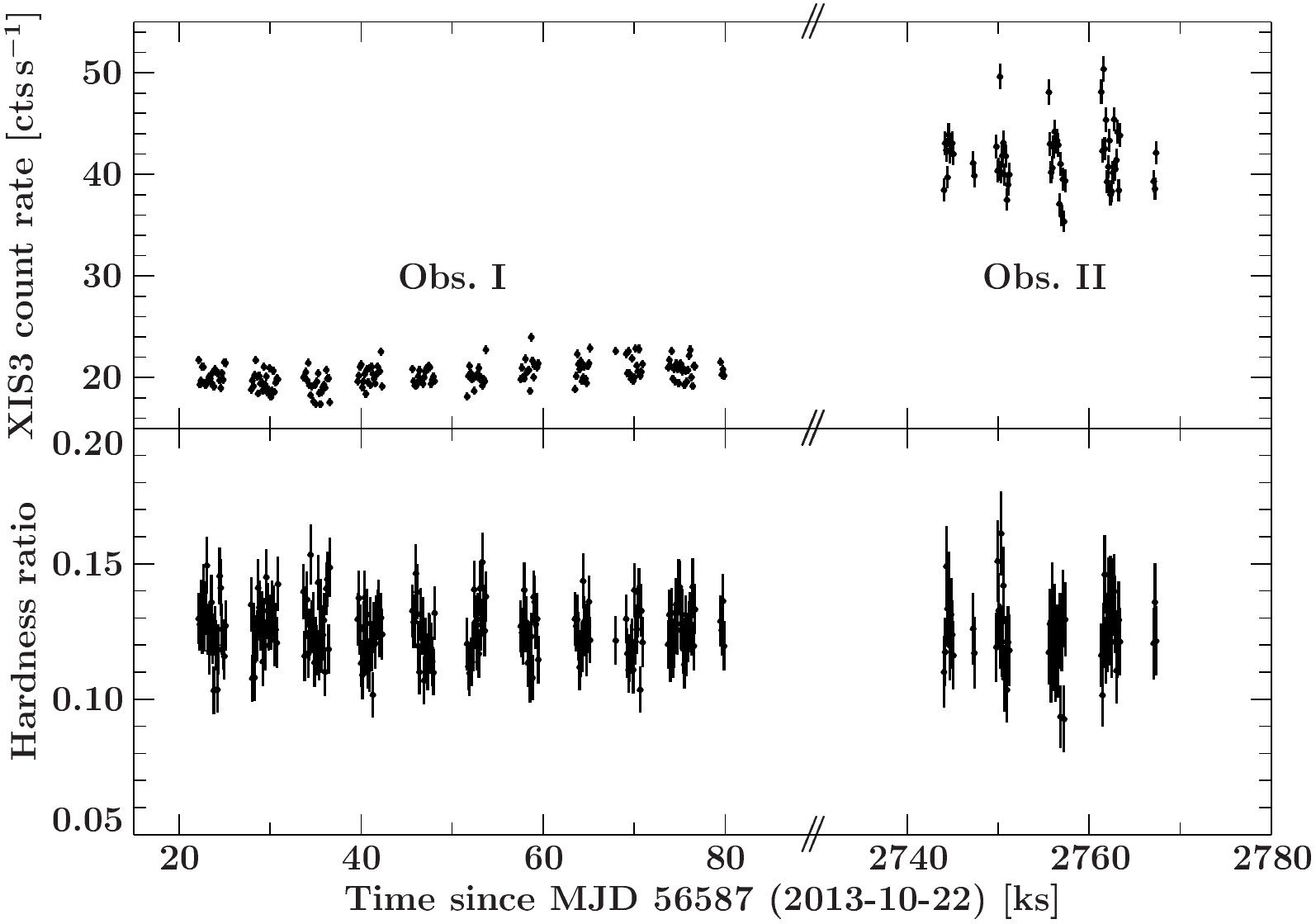}}
 \caption{Lightcurve and hardness ratio of XIS3 of both observations
   with 128\,s time bins. The lightcurve covers the energy range
   1--10\,keV. The hardness ratio is defined as the ratio of the count
   rate in the 7--10\,keV band to the count rate in the 1--4\,keV
   band.}
 \label{fig:xis_lightcurve}
\end{figure}

The lightcurves of both \suz observations show no sign of flaring or
significant variability and therefore indicate that time resolved
spectroscopy is not necessary. The absolute variability amplitude in
the lightcurve is higher in Obs.~II, but the relative variability is
similar. We note that the hardness ratio was roughly the same in both
observations (see Fig.~\ref{fig:xis_lightcurve}).

\begin{figure}
 \resizebox{\hsize}{!}{\includegraphics{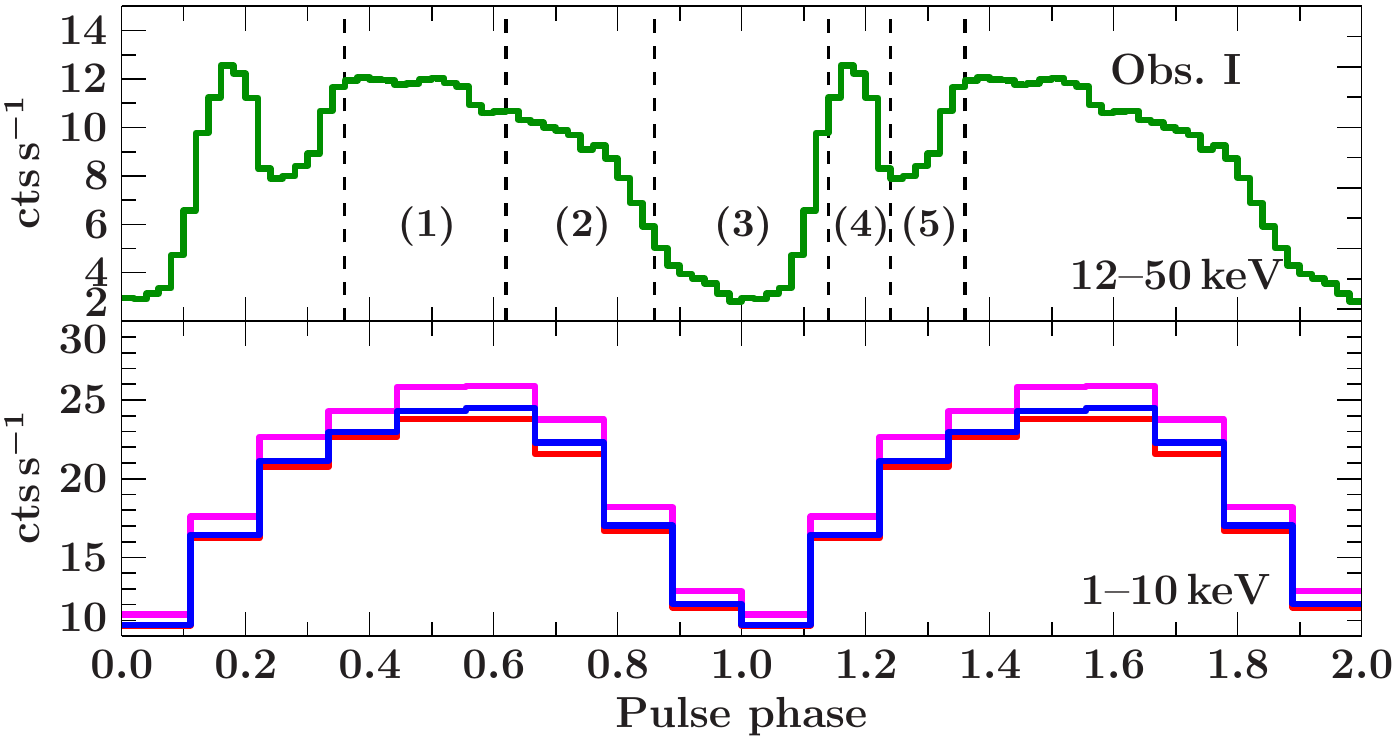}}
 \caption{Background-subtracted pulse profiles of Obs.~I. The upper
   panel shows the PIN profiles with the phase intervals chosen for
   phase-resolved spectral analysis. The lower panel shows the XIS0
   (blue), XIS1 (red), XIS3 (magenta) profiles. The pulse profiles are
   very similar to those of Obs.~II. All profiles are shown twice for
   clarity.}
 \label{fig:pulseprofiles}
\end{figure}

\begin{figure}
 \resizebox{\hsize}{!}{\includegraphics{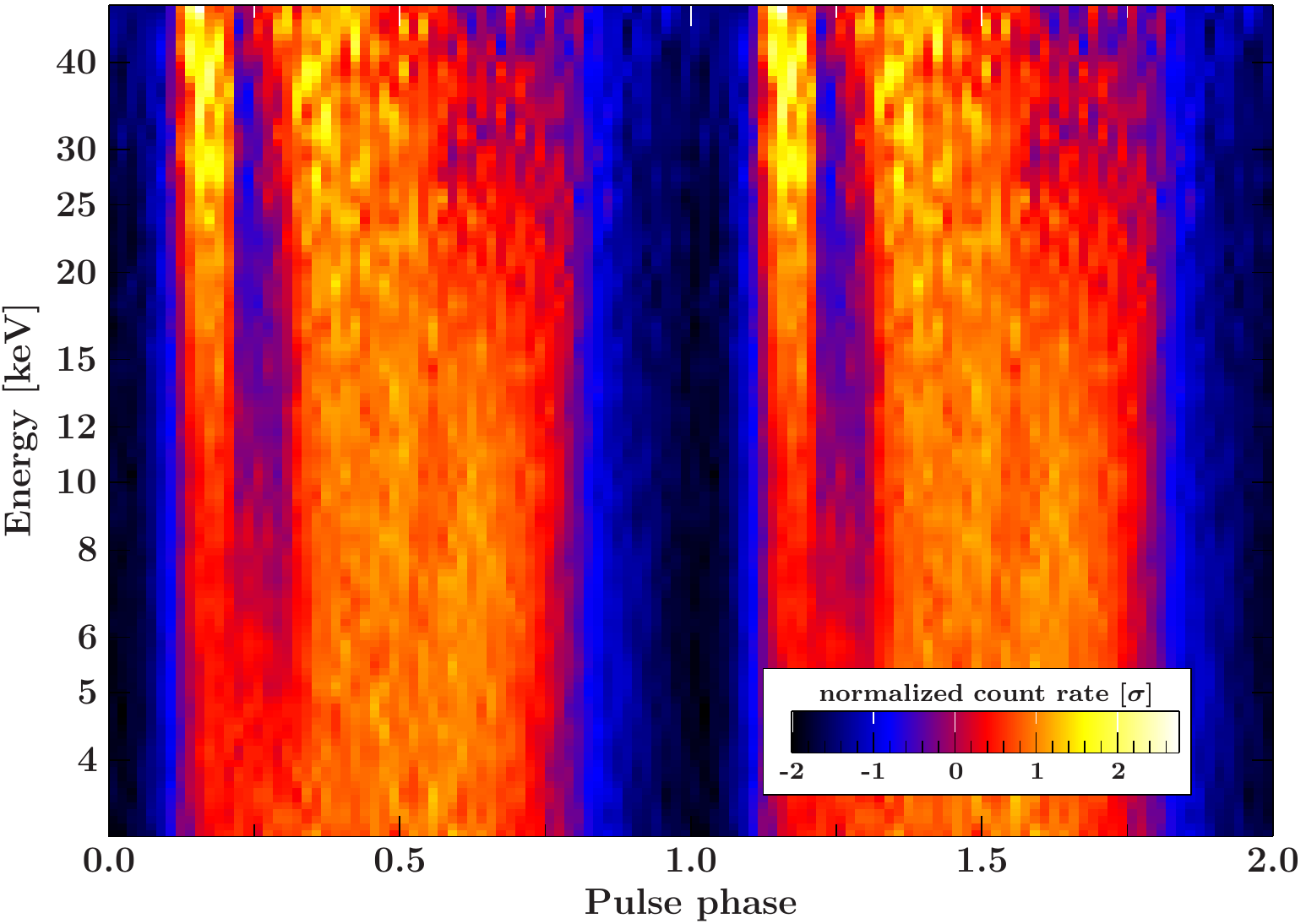}}
 \caption{Color-coded map of the count rate distribution of the
   \nustar observation as a function of energy and pulse phase. Each
   row represents a normalized pulse profile at the respective energy
   and each column, technically, a phase-resolved spectrum. All
   profiles are shown twice for clarity.}
 \label{fig:pulsemap}
\end{figure}

Pulse periods were determined for both observations using the epoch
folding technique \citep{Leahy1983}: Pulse profiles are calculated for
a series of test periods, assuming that for the correct period the
corresponding pulse profile shows the most distinct shape, whereas for
any other test period the averaging over a large number of pulses
leads to a smoothing of the resulting profile. We did not use GSO for
the pulse period determination because of its low statistics.
Uncertainties of the pulse periods were estimated by simulating
lightcurves based on the previously determined pulse profiles with
additional Poisson noise. Epoch folding was then applied to a large
number (in our case 10\,000) of simulated lightcurves. The standard
deviation of the pulse periods obtained with this simulation was taken
as uncertainty of the pulse period of the data. All \suz detectors are
in very good agreement with each other, but we quote the value of PIN,
since it has the highest time resolution. The measured pulse periods
are 18.80876(7)\,s and 18.78896(7)\,s for Obs.~I and Obs.~II,
respectively. Thus, the source showed spin-up between the observations
as was also observed by \citet{Furst2014}. We note, however, that the
lightcurves have not been corrected for binary motion because of an
accumulation of the uncertainties of the orbital parameters since
2000/2001. We estimated the possible impact of binary motion on the
determined spin-up by correcting the photon arrival times using the
orbital parameters of \citet{Galloway2004}. We found that, even
assuming that the phase information is lost completely, binary motion
can only account for $\sim$30\% of the observed pulse period
change. We therefore conclude that the observed spin-up is at least
partly caused by the transfer of angular momentum to the neutron star.

The shape and evolution of the pulse profile look very similar for
both observations and are therefore shown only for the first
observation in Fig.~\ref{fig:pulseprofiles}. The pulse profiles in the
XIS and PIN are significantly different from each other. The soft
pulse profile consists of only one broad peak, whereas the hard
profile is more complex and consists of a narrow peak and a broad peak
at higher energies. This characteristic evolution of the pulse profile
with energy has already been observed during previous outbursts
\citep[see, e.g.,][]{Tsygankov2005, Naik2006}.

Using the \suz data alone, however, the analysis of evolution of the
pulse profile at low energies is limited by the 2\,s time resolution
of the XIS. We took advantage of the excellent time resolution of
\nustar to investigate the pulse profile evolution also at soft
energies in more detail. We determined the local pulse period of the
\nustar data to be 18.78779\,s. We extracted light curves in 40 energy
bins between 3--50\,keV and folded them on this pulse period using 60
phase bins. Figure~\ref{fig:pulsemap} shows the resulting count rate
distribution of \nustar as a function of energy and pulse phase. To
account for the high-energy roll-over of the spectrum of the pulsar,
all pulse profiles were normalized to their mean count rate and are
given in units of their standard deviation. The pulse map shows that
the secondary peak starts to form at $\sim$6\,keV and is of comparable
strength to the main peak above $\sim$15\,keV. Interestingly, we do
not find any significant phase shifts around the cyclotron line energy
as predicted by theory \citep{Schoenherr2014} and as observed in
\object{4U\,0115+634} by \citet{Ferrigno2011}.
 
\section{Phase-averaged spectroscopy}\label{sec:phase_averaged}
We now turn to a description of the X-ray spectrum of the source in
both observations. For the first observation, the XIS spectra were
jointly re-binned to a minimum signal to noise ratio (\snr) of 80,
except for the energy range of 6--7\,keV, which was re-binned to a
minimum \snr of 65 to ensure higher resolution in the iron line
region. The intrinsic energy resolution of XIS, which is around
190\,eV FWHM at $\sim$6\,keV \citep{Ozawa2009}, is oversampled by a
factor of 4--5 in the iron line region. The PIN spectrum was re-binned
to a minimum \snr of 25 and for GSO a channel binning factor of three
was applied. For the second observation we chose a minimum \snr of 70
for XIS (50 for 6--7\,keV) to account for the shorter exposure time
but re-binned PIN and GSO with the requirements of the first
observation.

\begin{figure}
\resizebox{\hsize}{!}{\includegraphics{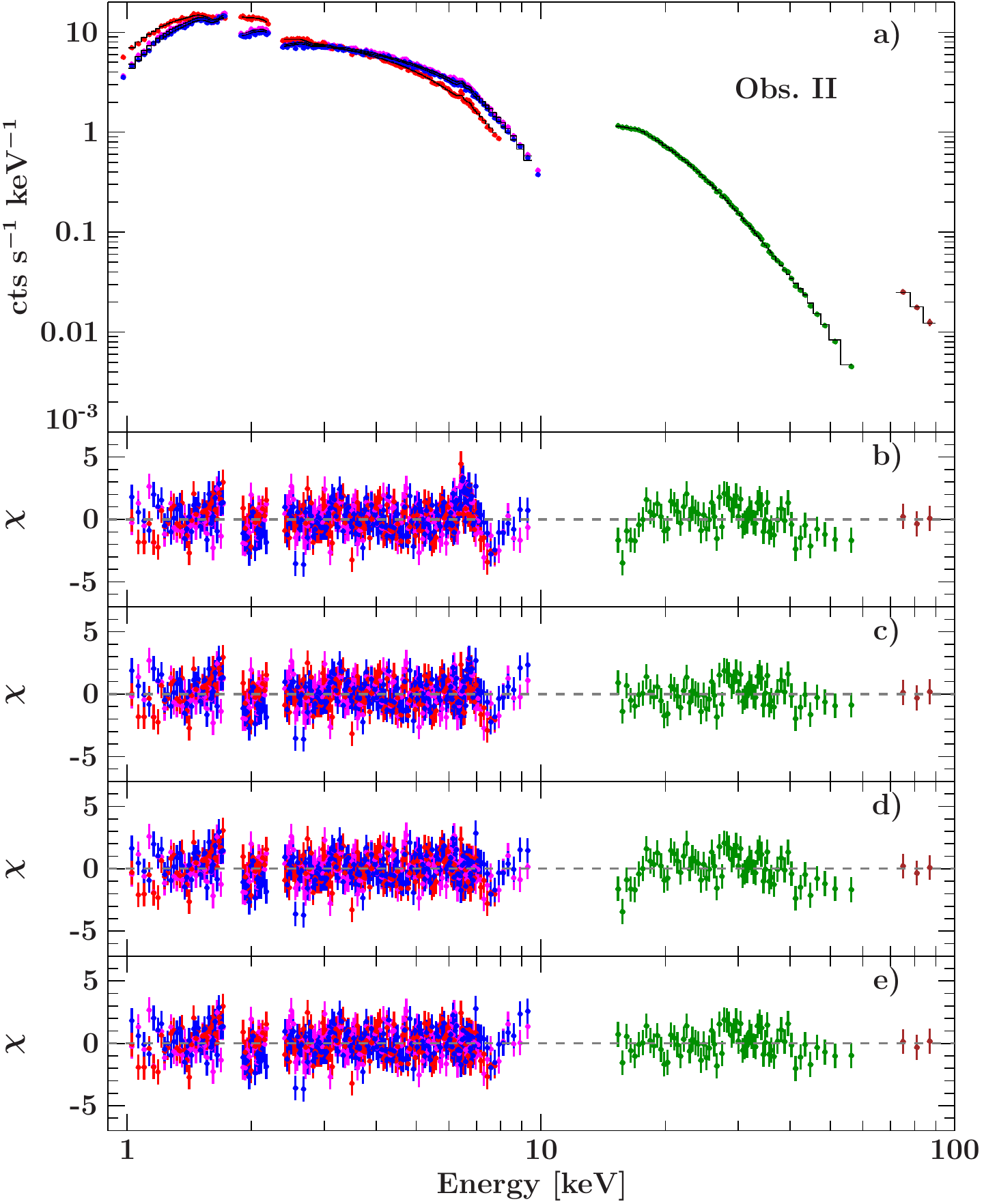}}
\caption{Panel \textbf{a)} Phase-averaged spectrum of Obs.~II with
  best-fit model. XIS0 (blue), XIS1 (red), XIS3 (magenta), PIN (green)
  and GSO (brown). \textbf{b)} Residuals of continuum model without
  CRSF and Fe lines. \textbf{c)} Residuals of continuum model with
  narrow K$\alpha$ and K$\beta$ lines of neutral iron only and CRSF.
  \textbf{d)} Residuals of continuum model with both neutral and
  He-like Fe lines but without CRSF. \textbf{e)} \ Residuals of
  best-fit model including CRSF and Fe-lines.}
\label{fig:sectrum_phaseav_obs2}
\end{figure}

Figure~\ref{fig:sectrum_phaseav_obs2} shows the phase-averaged
spectrum with the best-fit model for the second observation and the
best fit parameters for both observations are given in
Table~\ref{tab:phaseav_fit_par}.

The best fitting continuum model is an absorbed power law with an
exponential cut-off of the form
\begin{equation}
 \texttt{CutoffPL}(E)\propto E^{-\Gamma}\exp(-E/E_\mathrm{fold})~,
\end{equation}
as used in ISIS/XSPEC with the photon index $\Gamma$ and the folding
energy $E_\mathrm{fold}$. An additional black body component with
temperature $kT_\mathrm{BB}$ was required to describe the spectrum
below $\sim$10\,keV. Following \citet{Furst2014}, we used an updated
version of the absorption model \texttt{tbabs}\footnote{see
  \url{http://pulsar.sternwarte.uni-erlangen.de/wilms/research/tbabs/}},
called \texttt{tbnew} with abundances and cross sections set according
to \citet{Wilms2000} and \citet{Verner1996}, respectively.

\citet{Tsujimoto2011} systematically investigated the
cross-calibration of the individual XIS and reported that spectra
observed with XIS1 tend to be slightly softer compared to XIS0 and
XIS3. We therefore fitted the photon index of the back-illuminated
XIS1 separately from the other XIS and HXD. The normalization and
folding energy were required to be the same for all detectors during
the fit. We confirm the result of \citet{Tsujimoto2011} that the
photon index obtained from XIS1 is slightly higher compared to the
other XIS and HXD, which agree with each other to within the
confidence levels. Note that due to the lower signal to noise ratio of
the phase-resolved spectra this deviation of XIS1 and the other XIS'
can be ignored in the phase-resolved analysis.

\begin{figure}
 \resizebox{\hsize}{!}{\includegraphics{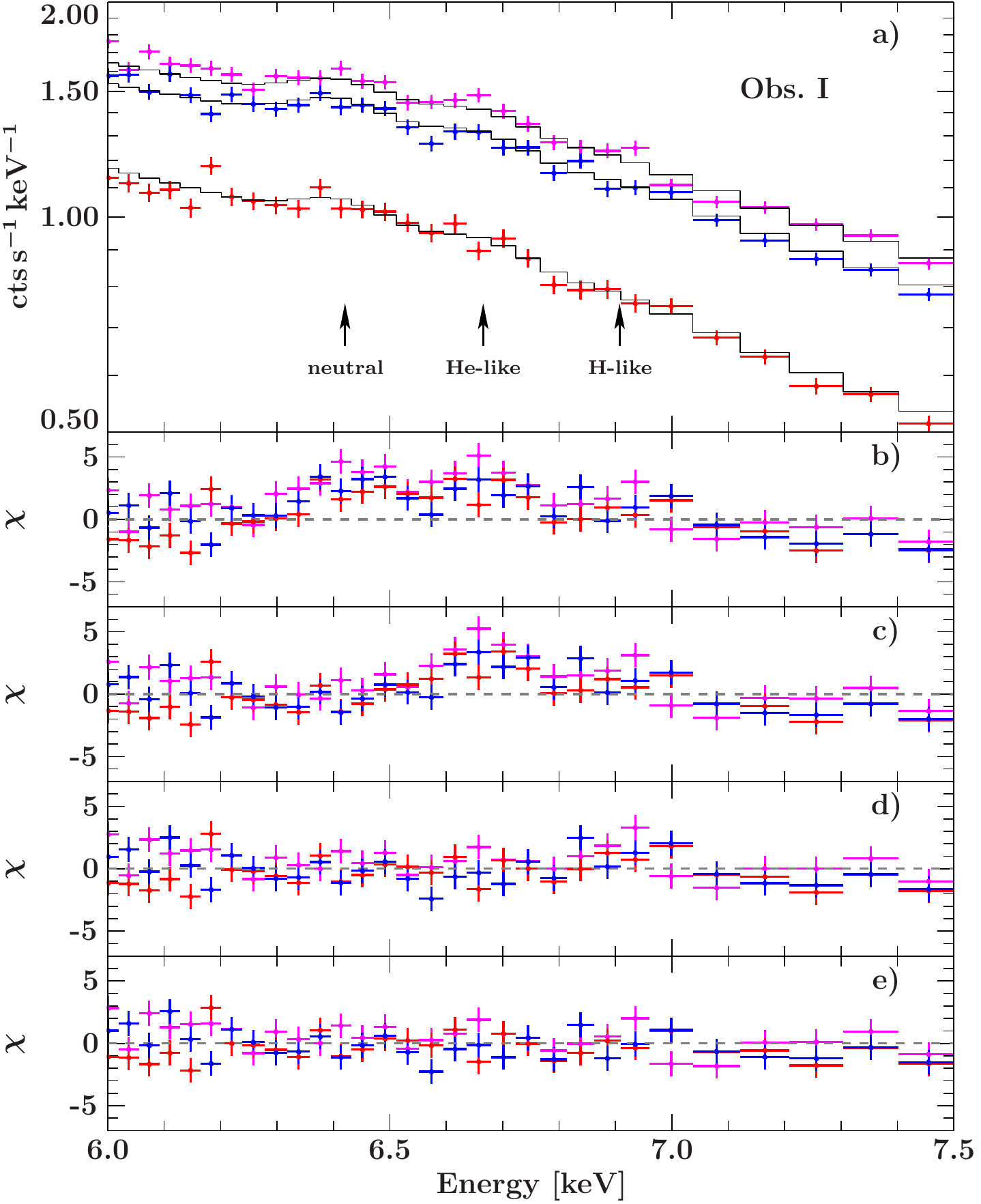}}
 \caption{Close-up plot of the iron line region for Obs.~I. Panel
   \textbf{a)} shows XIS0 (blue), XIS1 (red) and XIS3 (magenta) data
   with the best fitting model (black). Panel \textbf{b)} shows the
   residuals for the continuum model only, i.e., without any emission
   lines. Panel \textbf{c)} shows the residuals for the continuum
   model with the $\mathrm{K}\alpha$ line of neutral iron and panel
   \textbf{d)} shows the residuals for the continuum model with both,
   neutral and He-like, $\mathrm{K}\alpha$ lines. Panel \textbf{e)}
   shows the residuals for the continuum model with lines from
   neutral, He-like, and H-like iron included. All panels show the
   best fit for the respective model.}
 \label{fig:fe_lines}
\end{figure}

Figure~\ref{fig:fe_lines} shows the Fe line region of the spectrum for
Obs.~I and residuals for including different model components. For
both observations, the Fe~$\mathrm{K}\alpha$ and K$\beta$ lines were
modelled by two Gaussian emission lines. Furthermore, we fixed all
line widths to $10^{-6}\,\mathrm{keV}$, assuming that line broadening
is only due to the detector response. The Fe~K$\beta$ line is not
significantly detected, i.e., adding it does not significantly improve
the fit quality. However, if the line is due to fluorescence, then Fe
K$\beta$ emission has to be present at a known flux level and energy.
We therefore included such a line at 0.65\,keV higher than the
Fe~K$\alpha$ line energy with a flux of 13\% of the K$\alpha$ line to
the model \citep{Palmeri2003}. Note that this did not introduce
further fit parameters. Adding the K$\alpha$ and K$\beta$ line of
neutral iron to our final continuum model reduced the
$\chi^2/\mathrm{d.o.f}$ from 1178.11/718 to 1016.06/716 for Obs~I. and
from 878.14/565 to 790.95/563 for Obs.~II. It was furthermore
necessary to add a third emission line at $\sim$6.7\,keV, the
K$\alpha$ line energy of He-like iron. Including this component again
led to an improvement of $\chi^2/\mathrm{d.o.f}$ to 906.22/714 for
Obs.~I and 741.03/561 for Obs.~II. For the first observation, a fourth
narrow emission line at $\sim$6.9\,keV, the K$\alpha$ line energy of
H-like iron, was also required, reducing the $\chi^2/\mathrm{d.o.f}$
further to 893.53/712. This feature was not significantly detected in
the second observation, possibly because of the lower exposure. The
upper limit for the flux of this feature, however, is consistent with
that measured during the first observation. We determined the
significance of the H-like iron line with a Monte-Carlo approach
\citep[see, e.g.,][]{Protassov2002} and obtained a significance of
99.993\% with $10^5$ trials. For all other iron lines we found a
significance of greater than 99.999\%. Including a systematic
uncertainty of 1.3\% slightly reduces the detection significance of
the H-like iron line to 99.978\%. The other lines are not affected.

As mentioned above, \citet{Furst2014} discovered a cyclotron line at
12.2\,keV in the \nustar data. Unfortunately, this is exactly in the
gap between the XIS and the HXD spectra. Since the line is rather
wide, however, it can in principle still influence the \suz analysis.
We therefore applied the continuum model both with and without the
cyclotron line, describing the line by a Gaussian optical depth
profile

\begin{equation}
\texttt{gabs}(E) = \exp\left(\left(-
\frac{d_\mathrm{CRSF}}{\sqrt{2\pi}\sigma_\mathrm{CRSF}} \right)
\exp\left( -\frac{(E- E_\mathrm{CRSF})^2}{2\sigma_\mathrm{CRSF}^2}
\right)\right), 
\end{equation}
with the line energy $E_\mathrm{CRSF}$, the width
$\sigma_\mathrm{CRSF}$, and the depth $d_\mathrm{CRSF}$. 

As the \suz instruments do not cover the cyclotron line centroid
energy, not all three parameters can be constrained simultaneously. We
fixed the line energy and width to the values from \citet{Furst2014}
and only fitted the line strength. Fits that include the CRSF describe
the data better, especially for Obs.~II. In fact, \citet{Furst2014}
found the CRSF to be most prominent in the \nustar observation which
overlaps with the second \suz observation. Ignoring the CRSF leads to
negative residuals near 15\,keV (see
Fig.~\ref{fig:sectrum_phaseav_obs2}b and~d for Obs.~II). We caution
that this is by no means a detection or a confirmation of a CRSF in
the \suz data. However, there is a qualitative agreement of the \suz
data and the results by \citet{Furst2014} regarding the CRSF. Although
observed in a number of other sources \citep[e.g.,
\object{4U\,0115+634} or
\object{V\,0332+53},][respectively]{Heindl1999, Pottschmidt2005}, we
find no indications of a second harmonic, which we would expect around
24\,keV, although PIN and \nustar are both very sensitive in this
energy range. We determined the upper limits for the optical depth of
a second harmonic at 24.4\,keV assuming a width of 2.5\,keV and find
$d_\mathrm{CRSF} \le0.05$ for Obs.~I and $d_\mathrm{CRSF} \le0.11$ for
Obs.~II. Using the \nustar spectrum with exactly the same binning and
energy range restriction criteria as described by \citet{Furst2014} we
obtained $d_\mathrm{CRSF} \le0.10$. There are, however, also other
examples where harmonic lines have not yet been detected \citep[e.g.,
\object{Cen~X-3} or
\object{RX~J0520.5$-$6932},][respectively]{Suchy2008, Tendulkar2014}
although their energies are in principle accessible with current
instruments.

The best fit parameters are given in Table~\ref{tab:phaseav_fit_par}
and Fig.~\ref{fig:sectrum_phaseav_obs2} shows the phase-averaged
spectrum and best-fit model of the second observation.

The reduced $\chi^2$ values of our best fits are still
somewhat larger than 1. We ascribe this to calibration uncertainties
and potentially minor shortcomings of the source modeling. We tried
adding a systematic error in quadrature to the data and found an
assumed systematic error of 1.3\% to result in reduced
$\chi^2$ values of 0.93 for Obs.~I and 1.06 for Obs.~II for
the same degrees of freedom as given in
Table~\ref{tab:phaseav_fit_par}. However, the choice of the systematic
uncertainty is rather arbitrary and not explicitly recommended by any
dedicated calibration studies. We therefore do not include a
systematic uncertainty in our analysis.

\begin{table}
\caption{Best fit parameters and statistics for the phase-averaged spectra.}
\label{tab:phaseav_fit_par}
\centering
\begin{tabular}{lcc}
\hline\hline
Parameter & Obs.~I & Obs.~II\\
\hline
 $N_\mathrm{H}~[10^{22}\,\mathrm{cm}^{-2}]$ & $0.796\pm0.021$ & $0.784\pm0.027$ \\
 $\mathcal{F}_{15-50\,\mathrm{keV}}$\tablefootmark{a} & $3.70\pm0.02$ & $4.88\pm0.3$ \\
 $\Gamma_\mathrm{XIS0,3}$ & $0.97\pm0.04$ & $0.89\pm0.04$ \\
 $\Gamma_\mathrm{XIS1}$ & $1.00\pm0.04$ &  $0.95\pm0.04$ \\
 $\mathcal{F}_\mathrm{BB}$\tablefootmark{b} &$0.306\pm0.018$ & $0.479^{+0.029}_{-0.031}$ \\
 $kT_\mathrm{BB}~[\mathrm{keV}]$ & $0.581\pm0.010$ & $0.673\pm0.012$ \\
 $E_\mathrm{fold}~[\mathrm{keV}]$ & $23.3\pm0.9$ & $22.8\pm0.9$ \\
 $E_{\ion{Fe}{i}\,\mathrm{K}\alpha}~[\mathrm{keV}]$ & $6.420^{+0.020}_{-0.017}$ & $6.440^{+0.023}_{-0.017}$ \\
 $A_{\ion{Fe}{i}\,\mathrm{K}\alpha}$\tablefootmark{c} & $6.3\pm1.0$ & $8.4\pm1.5$ \\
 $E_{\ion{Fe}{xxv}\,\mathrm{K}\alpha}~[\mathrm{keV}]$ & $6.666\pm0.021$ & $6.729^{+0.029}_{-0.027}$ \\
 $A_{\ion{Fe}{xxv}\,\mathrm{K}\alpha}$\tablefootmark{c} & $5.9\pm1.0$ & $6.5^{+1.6}_{-1.5}$ \\
 $E_{\ion{Fe}{xxvi}\,\mathrm{K}\alpha}~[\mathrm{keV}]$ & $6.91\pm0.05$ & -- \\
 $A_{\ion{Fe}{xxvi}\,\mathrm{K}\alpha}$\tablefootmark{c} & $2.1\pm1.0$ & -- \\
 $E_\mathrm{CRSF}$\tablefootmark{d}~$[\mathrm{keV}]$ & 12.2 & 12.2 \\
$d_\mathrm{CRSF}$ & $0.36^{+0.15}_{-0.14}$ &  $0.48\pm0.14$ \\
 $\sigma_\mathrm{CRSF}$\tablefootmark{d}~[keV] & 2.5 & 2.5 \\ 
 $C_\mathrm{XIS0}$\tablefootmark{e} & $0.764^{+0.014}_{-0.013}$ & $0.952^{+0.021}_{-0.020}$ \\
 $C_\mathrm{XIS1}$\tablefootmark{e} & $0.738^{+0.019}_{-0.018}$ & $0.893^{+0.027}_{-0.026}$ \\
 $C_\mathrm{XIS3}$\tablefootmark{e} & $0.743\pm0.013$ & $0.967\pm0.021$ \\
 $C_\mathrm{PIN}$\tablefootmark{e} & $1$ & $1$ \\
 $C_\mathrm{GSO}$\tablefootmark{e} & $0.80\pm0.08$ & $0.79\pm0.06$ \\
\hline
 $\chi^2_\mathrm{red}(\mathrm{d.o.f})$ & 1.25~(712) & 1.32~(561)\\
\hline
\end{tabular}
\tablefoot{
  \tablefoottext{a}{Unabsorbed flux in the 15--50\,keV band in units of $10^{-9}\,\mathrm{erg}\,\mathrm{s}^{-1}\,\mathrm{cm}^{-2}$.}
  \tablefoottext{b}{Black body flux in the 1--10\,keV band in units of $10^{-9}\,\mathrm{erg}\,\mathrm{s}^{-1}\,\mathrm{cm}^{-2}$.}
  \tablefoottext{c}{Photon flux in units of $10^{-4}\,\mathrm{photons}\,\mathrm{s}^{-1}\,\mathrm{cm}^{-2}$.}
  \tablefoottext{d}{Fixed to the value from \citet{Furst2014}.}
  \tablefoottext{e}{Detector flux cross-calibration constant, relative to PIN.}
}
\end{table} 

We note a strong difference for the XIS detector normalization
constants between both observations. We attribute this difference to
the different data extraction modes and the lacking attitude
correction for the first observation, as the simulation of the
effective area of XIS, which is performed by the FTOOL
\texttt{xissimarfgen} in the course of data reduction, is highly
sensitive to the source position and extent. Possible deviations of
the apparent source position due to thermal wobbling might therefore
introduce an unknown systematic uncertainty to the flux measured by
XIS. For this reason, we normalized the detector constants with
respect to PIN. The fitted fluxes in the 15--50\,keV band are about
10\% higher than the corresponding \nustar observations, which agrees
with flux calibration uncertainties \citep{Madsen2015}.

For future reference, we also determined the continuum flux for other
energy bands than given in Table~\ref{tab:phaseav_fit_par}. For the
first observation, the measured fluxes are
$(2.46\pm0.05)\times10^{-9}\,\mathrm{erg}\,\mathrm{s}^{-1}\,\mathrm{cm}^{-2}$
and
$(7.23\pm0.07)\times10^{-9}\,\mathrm{erg}\,\mathrm{s}^{-1}\,\mathrm{cm}^{-2}$
in the 2--10\,keV and 3--60\,keV band, respectively, and for the
second observation
$3.05^{+0.10}_{-0.09}\times10^{-9}\,\mathrm{erg}\,\mathrm{s}^{-1}\,\mathrm{cm}^{-2}$
and
$(9.32\pm0.10)\times10^{-9}\,\mathrm{erg}\,\mathrm{s}^{-1}\,\mathrm{cm}^{-2}$
for the corresponding energy bands.

\begin{figure}
 \resizebox{\hsize}{!}{\includegraphics{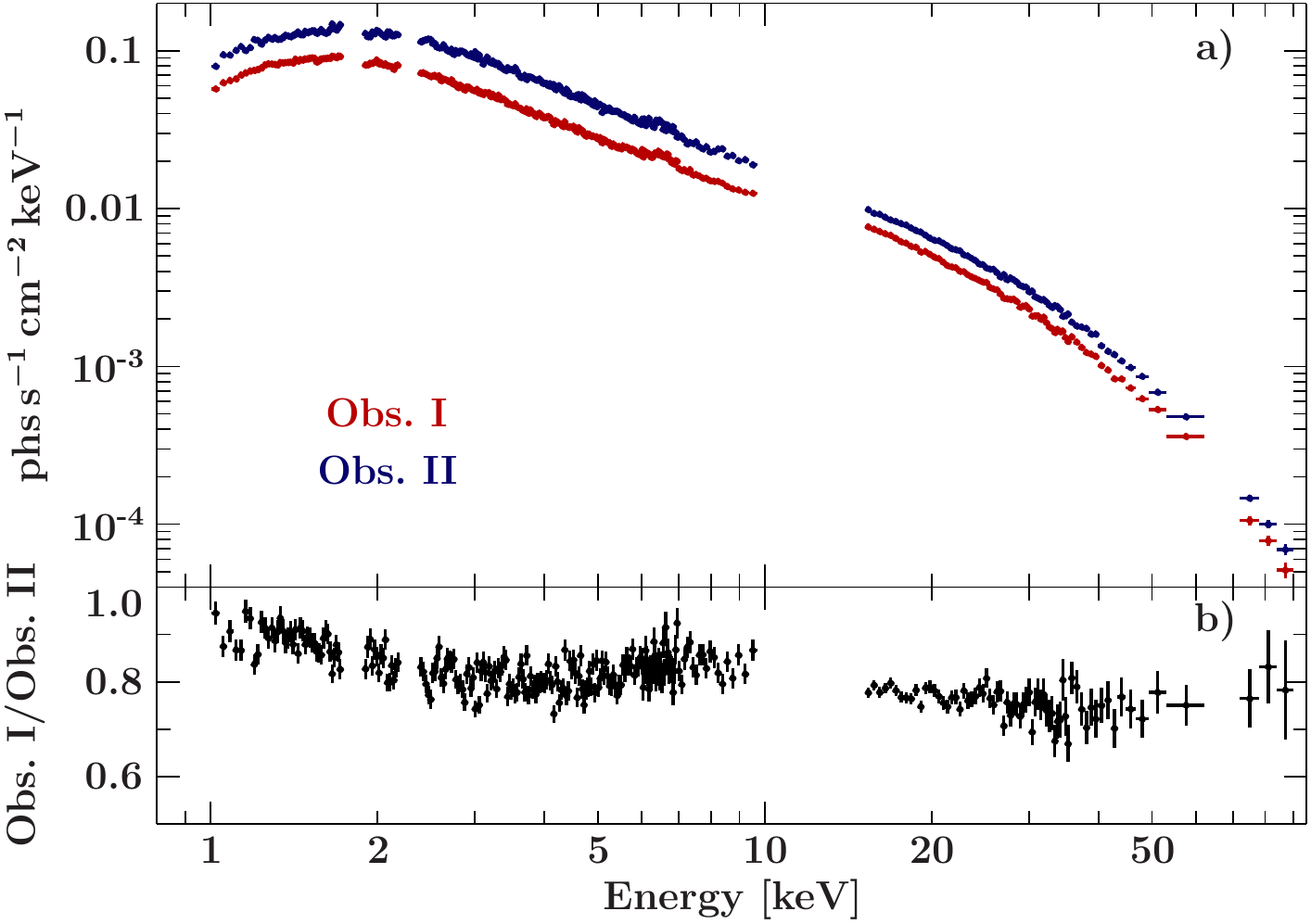}}
\caption{\textbf{a)} Unfolded phase-averaged spectra of Obs.~I (red)
  and Obs.~II (blue) and \textbf{b)} ratio of the phase-averaged
  spectra of Obs.~I and Obs.~II. For clarity, only XIS3, PIN and GSO
  data are shown in both panels. The ratios of the individual
  detectors have been corrected for the respective cross-calibration
  constants given in Table~\ref{tab:phaseav_fit_par}.}
 \label{fig:spec_ratio}
\end{figure}

Figure~\ref{fig:spec_ratio} shows the unfolded spectra of Obs.~I and
Obs.~II in comparison and the photon ratio of both observations. We
observe a slight change in spectral slope below $\sim$2.5\,keV, while
the ratio is mainly flat above this energy, with slight differences in
the Fe band. Above the cyclotron line energy the spectra are very
similar in shape. Comparing the fit parameters of both \suz
observations, we find a decrease of the photon index and the folding
energies towards higher flux. \citet{Furst2014} observed the opposite
trend regarding the first and second \nustar observation. There is,
however, an artificial correlation between these two parameters, which
is not taken into account here, so we cannot conclude that the \suz
and \nustar results contradict each other. The excess of soft photons
in Obs.~I below $\sim$2.5\,keV is at least partly explained by the
increase of the black body temperature from Obs.~I to Obs.~II, while
the relative normalization of the black body component is very similar
in both observations. This behavior is consistent with \nustar. See
\citet{Furst2014} for a more detailed investigation of the parameter
evolution over the outburst.

The fitted $N_\mathrm{H}$ value is comparable to the value of
$0.89\times10^{22}\,\mathrm{cm}^{-2}$ given by The
Leiden/Argentine/Bonn (LAB) Survey of Galactic HI
\citep{Kalberla2005}.

\section{Phase-resolved spectroscopy}\label{sec:phase_resolved}

In order to study the variation of the continuum with the viewing
angle onto the neutron star, we extracted spectra for five individual
pulse phase intervals (Phase bin [Pb]~1--5, see
Fig.~\ref{fig:pulseprofiles} for the definition of the phase
intervals). The width and location of the phase intervals were chosen
to ensure sufficient \snr as well as to distinguish the most
characteristic features of the pulse profiles.  In order to account
for lower statistics due to splitting the observations into phasebins,
we reduced the required minimum \snr for re-binning the XIS spectra to
50 (45 for 6--7\,keV) for all phase intervals with exception of the
phase interval covering the pulse profile minimum (Pb~3) where we
required a minimum \snr of 40 (35 for 6--7\,keV). The PIN spectra were
re-binned to a minimum \snr of 15. We ignored the GSO spectra for the
phase-resolved analysis because of their low statistics.

\begin{figure}
 \resizebox{\hsize}{!}{\includegraphics{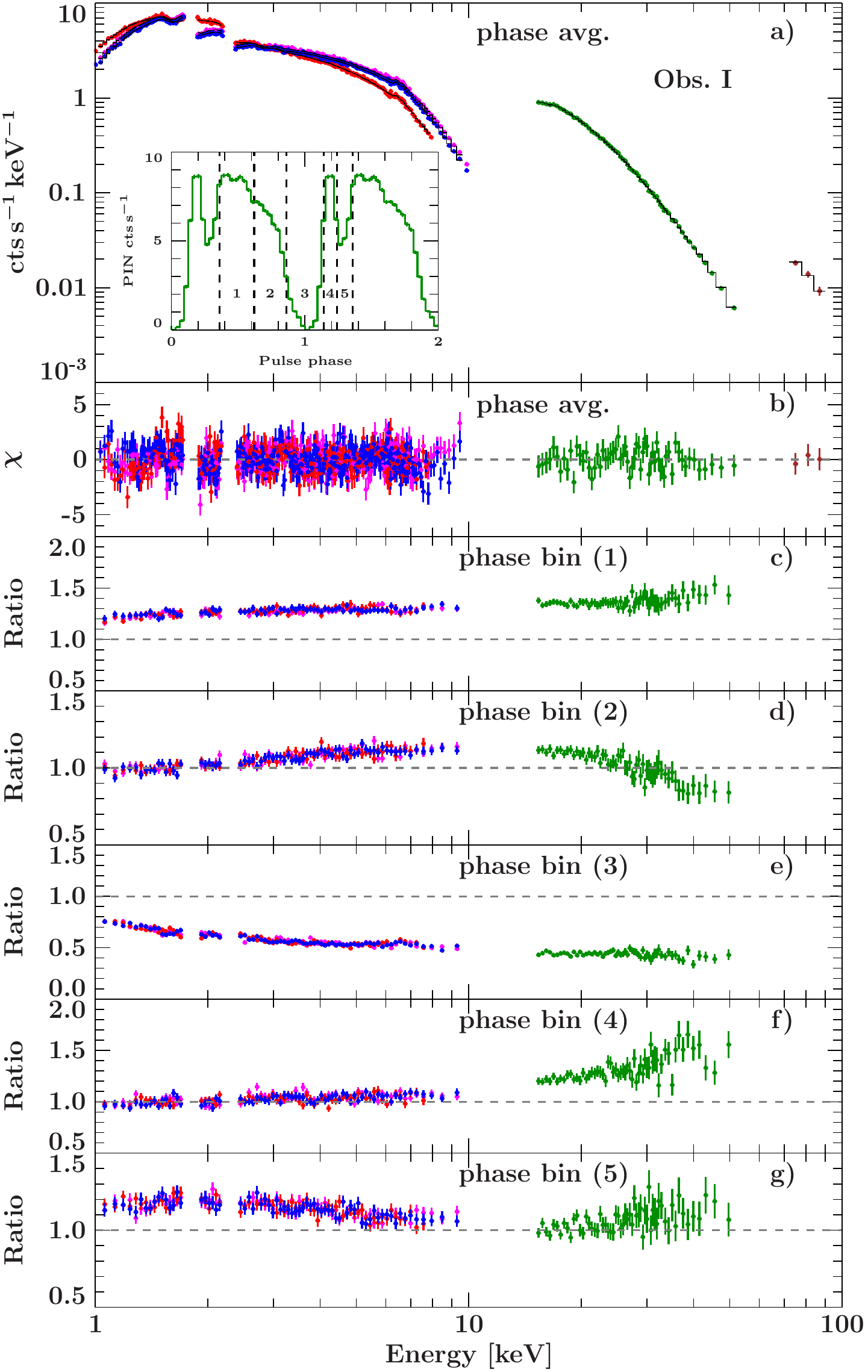}}
 \caption{Panels \textbf{a)} and \textbf{b)} show the phase-averaged
   spectrum of XIS0 (blue), XIS1 (red), XIS3 (magenta), PIN (green)
   and GSO (brown) of Obs.~I with best fit model and residuals. Panels
   \textbf{c)} through \textbf{g)} show the count rates of the respective
   phase-resolved spectra (1) to (5), divided by the count rate of the
   phase-averaged spectrum.}
 \label{fig:spectrum_ratios_obs1}
\end{figure}

We first calculated the ratio of all phase-resolved spectra with
respect to the phase-averaged spectrum, allowing us to investigate the
evolution of the phase-resolved spectra in a model-independent way
(Fig.~\ref{fig:spectrum_ratios_obs1}). The spectrum of Pb~1 is most
similar in shape to the phase-averaged spectrum, which is not
surprising as this phase interval covers the peak of the main pulse
and therefore contributes heavily to the phase-averaged spectrum. All
other ratios indicate changes in spectral shape. The spectrum of the
main minimum (Pb~3) is softer than that of the main pulse and the
spectrum of the emerging peak (Pb~4) is harder. The differences of the
main pulse spectrum with respect to the spectra of the declining main
pulse (Pb~2) and the emerging minimum (Pb~5) are more complex.

We used the same spectral model for the phase-resolved as for the
phase-averaged analysis, but without the H-like iron emission for
the first observation. Since we expect the detector cross-calibration
constants and the iron line energies to be independent of the viewing
angle onto the neutron star, these were constrained to be the same for
all phase intervals. On our final fits, these parameters ended up to be
identical to those of the phase-averaged analysis to within their
confidence intervals. All other parameters were fitted individually
for each phase interval.

\begin{figure}
 \resizebox{\hsize}{!}{\includegraphics{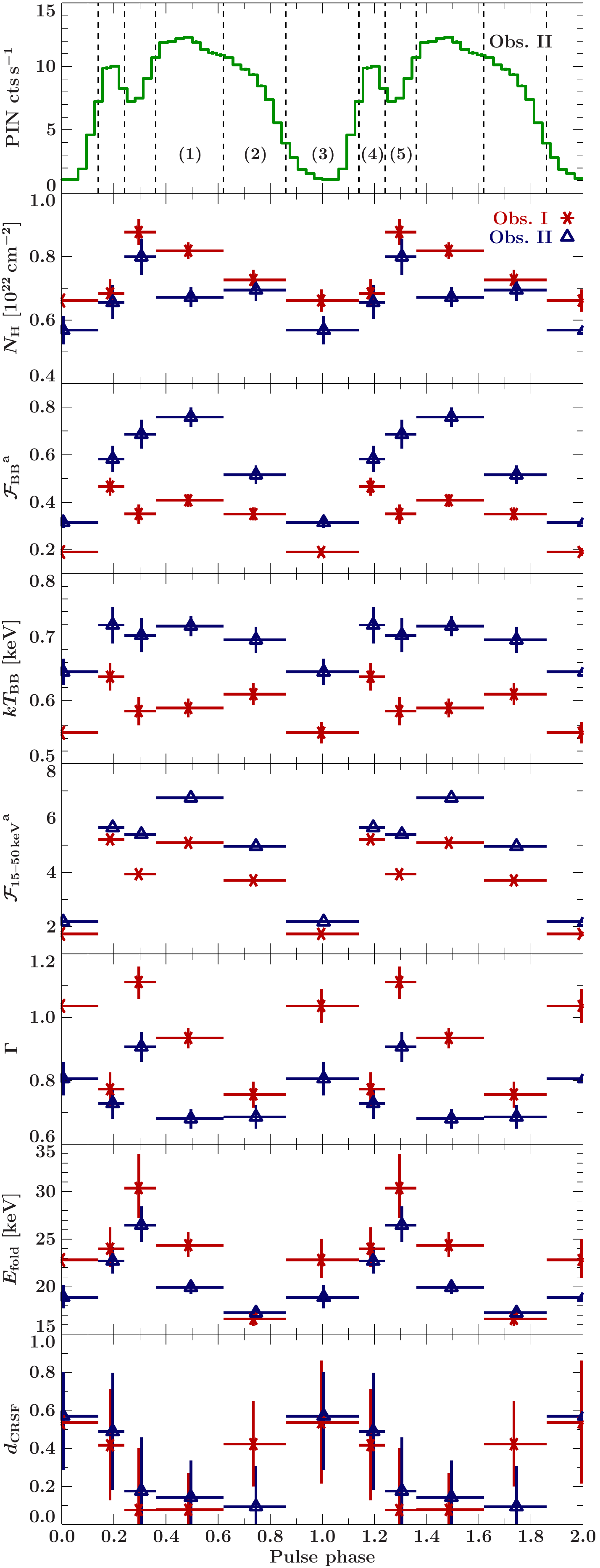}}
 \caption{Evolution of the best fit parameters of Obs.~I (red) and
   Obs.~II (blue) over pulse phase. The upper panel shows the PIN
   pulse profile of Obs.~II over its complete energy range. Dashed
   lines indicate the selected phase intervals. All fluxes
   ($^\text{a}$) are given in units of
   $10^{-9}\,\mathrm{erg}\,\mathrm{s}^{-1}\,\mathrm{cm}^{-2}$.
   Profiles and parameter evolutions are shown twice for clarity.}
 \label{fig:parameter_evolution}
\end{figure}

\begin{figure}
 \resizebox{\hsize}{!}{\includegraphics{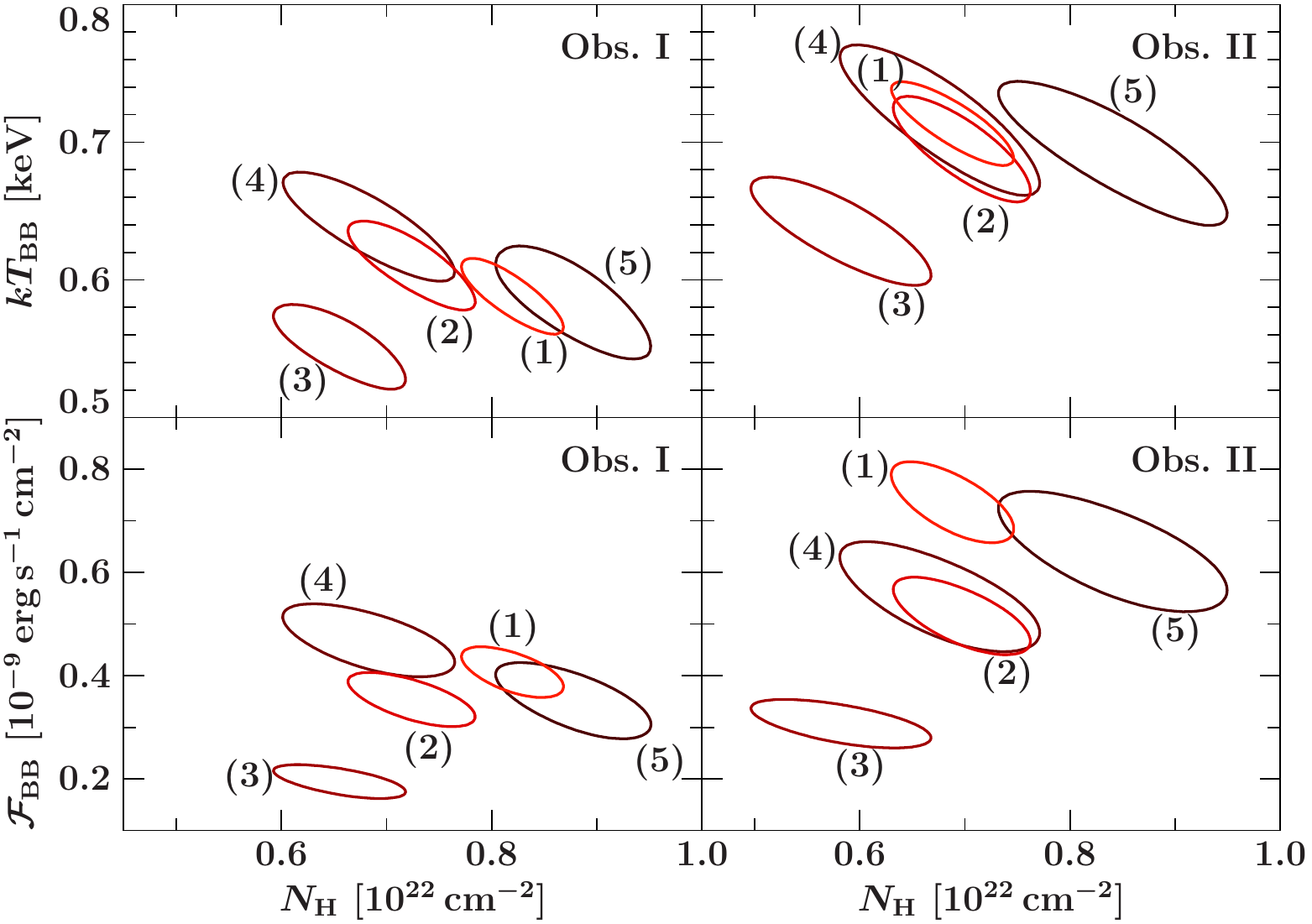}}

 \vspace*{0.5\baselineskip}

 \resizebox{\hsize}{!}{\includegraphics{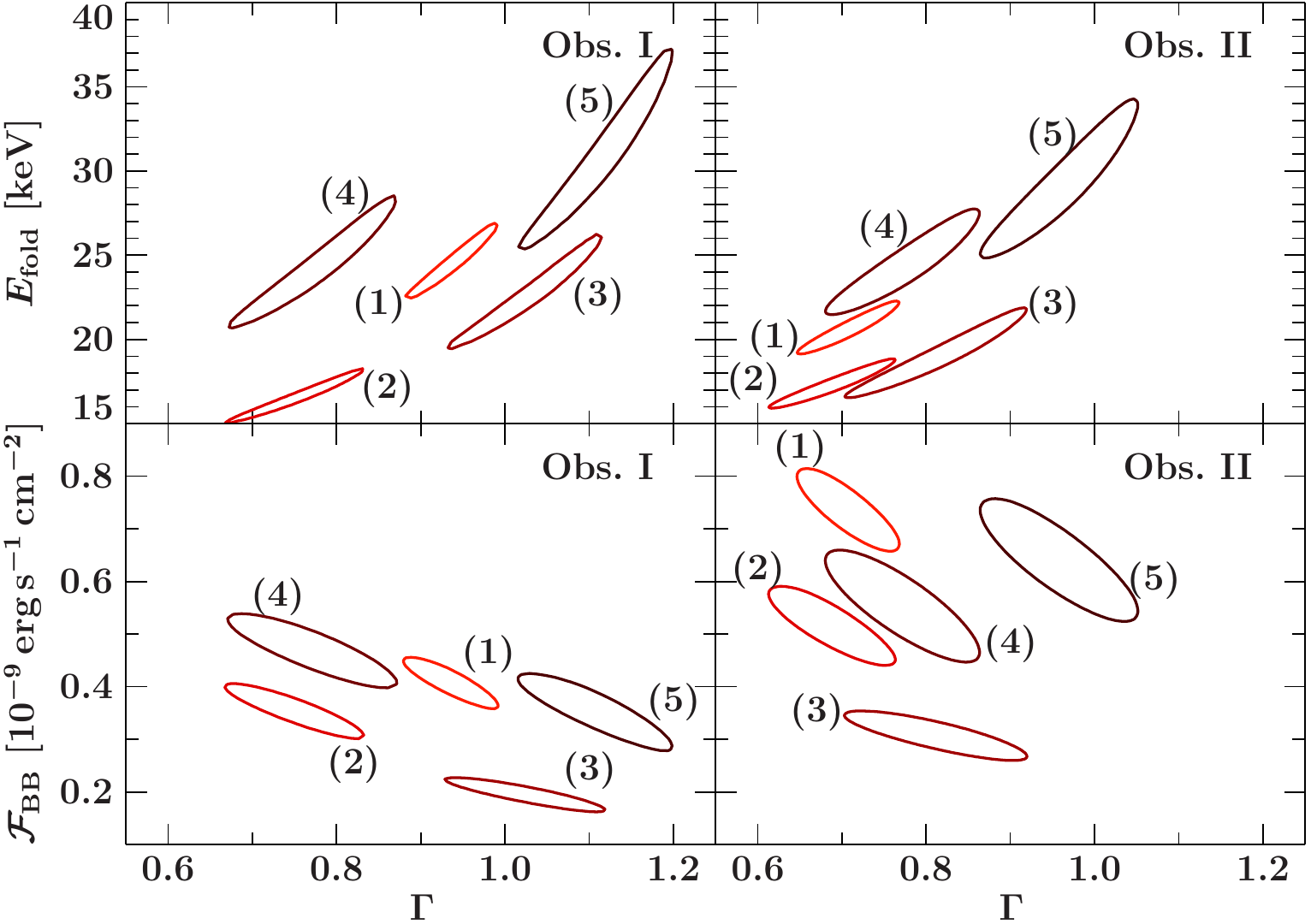}}
 \caption{Contour maps showing the $3\sigma$ confidence regions for
   the phase resolved spectra of both observations. \emph{Top:}
   Black body temperature and flux versus $N_\mathrm{H}$.
   \emph{Bottom:} Folding energy $E_\mathrm{fold}$ and black body
   temperature versus photon index $\Gamma$. Tied parameters (the iron
   line energies and the detector cross-normalization constants) were
   held constant for the calculation. The labels (1)--(5) indicate the
   respective phase interval. }\label{fig:contours}
\end{figure}

Figure~\ref{fig:parameter_evolution} presents the evolution of some of
the fit parameters over pulse phase for both observations. We observe
variations for all continuum parameters as well as of the CRSF
strength over pulse phase. Variations of the hard power law flux
$\mathcal{F}_{15-50\,\mathrm{keV}}$, the black body flux
$\mathcal{F}_\mathrm{BB}$, and the iron line flux (not shown) mostly
follow the pulse profile. Trends are generally the same for Obs.~I and
Obs.~II, while absolute parameter values can be moderately different.
In order to evaluate the significance of parameter differences in the
presence of possible artificial dependencies we calculated confidence
contours for two parameters of interest for several parameter pairs,
see Fig.~\ref{fig:contours}.

The emerging minimum (Pb~5) shows the highest absorption and the main
minimum (Pb~3) shows the lowest as well as a comparatively weak and
cool black body. These two phase bins generally display the most
distinct changes compared to their neighbors. The contour plots
confirm that the difference in $N_\mathrm{H}$ between the two extremes
is significant. Regarding the black body temperature they reveal that
-- with exception of the main minimum -- the phase intervals are
consistent with a $kT_\mathrm{BB}$ of $\sim$0.6\,keV and
$\sim$0.7\,keV for the first and second observation,
respectively\footnote{The individual $N_\mathrm{H}$-$kT_\mathrm{BB}$
  and $N_\mathrm{H}$-$\mathcal{F}_\mathrm{BB}$ contours reveal
  artificial parameter dependencies in the sense that higher values of
  $N_\mathrm{H}$ correspond to lower black body temperatures and
  fluxes, although at first glance one might expect the absorption
  component to ``compensate'' for emission in the soft range. But the
  individual $\Gamma$-$\mathcal{F}_\mathrm{BB}$ contours show that the
  trade-off happens with a softer power law instead.}. Changes of the
absorption column and black body temperature could not be constrained
by the $>3\,\mathrm{keV}$ phase-resolved \nustar analysis, where
$N_\mathrm{H}$ was fixed to $0.845\times10^{22}\,\mathrm{cm}^{-2}$ and
$kT_\mathrm{BB}$ to $\sim$0.6\,keV \citep{Furst2014}.

Confirming the picture obtained from the spectral ratios of
Fig.~\ref{fig:spectrum_ratios_obs1}, the phase variation of $\Gamma$
shows that the power law is comparatively soft during the main minimum
(Pb~3) and the emerging minimum (Pb~5). For Obs.~I, the contours show
that the softest and hardest $\Gamma$ values are indeed significantly
different from each other. The folding energy is roughly correlated
with $\Gamma$. The $E_\mathrm{fold}$-$\Gamma$ contours reveal that
this variation is at least in part due to the model intrinsic
correlation between these two parameters. Our results indicate that
$\Gamma$ and $E_\mathrm{fold}$ are high during the secondary minimum
(Pb~5) and that the CRSF is strongest during the main minimum (Pb~3)
and secondary peak (Pb~4) which is qualitatively consistent with the
higher resolution \nustar analysis \citep[20 phase
  bins,][]{Furst2014}. The \nustar analysis also found a slight
softening during the main minimum.

\section{Discussion and Conclusions}\label{sec:discussion}
In this paper we reported on spectral and timing analyses of two
observations of the 2013 outburst of \ks. The phase-averaged spectrum
shows emission features at energies $\sim$6.4\,keV, $\sim$6.7\,keV,
and -- for the first observation -- also at $\sim$6.9\,keV, which we
interpret as K$\alpha$ emission from neutral, He-like, and H-like
iron. This is the first time lines from different ionization states of
iron are detected in this source. The neutral line could also be a
blend of other ionization states, which cannot be distinguished with
the available data. 

Observations of the 2000/2001 outburst already found indications for
the presence of neutral \citep[][with \rxte]{Tsygankov2005} or He-like
iron \citep[][with \textsl{BeppoSAX}]{Naik2006}, however, none of these
observations showed the different ionization stages at the same time,
probably due to the lower energy resolution of the earlier
instruments. \nustar, with an energy resolution that is a factor of
$\sim$2 worse than the XIS, only detected one line feature at
6.5\,keV, hinting at the presence of ionization features
\citep{Furst2014}, similar to the variability of the line energy seen
with \rxte \citep{Tsygankov2005}. The accreting pulsar best known to
show emission lines from neutral, He-like and H-like iron is the
persistent source \object{Cen~X-3}, where the individual line strengths are
varying over the binary orbit, constraining different emission regions
\citep{Naik2011a}.

Although the $\sim$12.2\,keV center energy of the cyclotron line is
not covered by the \suz instruments, this feature was included in the
spectral model, as the line wings affect the spectral shape. The
correction due to the cyclotron line is most prominent during the
pulse minimum and the emerging second peak, consistent with the
\nustar-detected line variability \citep{Furst2014}. 

Consistent with earlier observations \citep[e.g.,][]{Naik2006}, the
pulse profiles are highly energy dependent, especially the evolution
from one broad peak to one broad and one narrow peak towards higher
energies is observed. Many neutron star binaries show the opposite
behavior with peaks vanishing at high energies, e.g.,
\object{4U\,0115+634} \citep{Mueller2010}, \object{4U\,1909+07}
\citep{Furst2011}, \object{Vela~X-1} \citep{Kreykenbohm2002},
\object{GX\,304$-$1} \citep{Devasia2011}, \object{1A\,1118$-$61}
\citep{Suchy2011, Maitra2012}, \object{GRO\,J1008$-$57}
\citep{Naik2011} or \object{EXO\,2030+375} \citep{Naik2013}. We note,
however, that \object{IGR\,J16393$-$4643} shows an evolution with
energy that is similar to that in \ks
\citep{Islam2015}. \citet{Islam2015} suggest that additional soft
photons from the off-pulse regions or a more direct view into the
emission region at the peak are the reason for the observed spectral
variation.

The characteristic narrow and sharp peak observed in the pulse
profiles of \ks could be caused by relativistic light bending effects,
allowing constraints on the inclination of the observer and the
magnetic field. Such parameters are generally very difficult to
determine. Any additional information on the geometry of the system,
as provided, e.g., by a sharp feature, reduces the number of free
parameters and thus simplifies the modeling of the pulse profiles. The
energy dependence might be caused by an energy-dependent beam pattern
and contributions from the accretion column, the neutron star surface
and a surrounding halo \citep{Kraus1989, Kraus2003, Falkner2013}. A
quantitative modeling of the pulse profiles is, however, beyond the
scope of this paper and will be presented in a forthcoming
publication.

The high sensitivity of \suz XIS in the soft X-rays allows us to
additionally investigate spectral variation over pulse phase below
3\,keV. We find the spectrum to be softer during the two minimum
phases Pb~3 and Pb~5. Despite strong correlations between the
continuum parameters (Fig.~\ref{fig:contours}), the soft spectrum
during Pb~3 appears to be due to a combination of a lower
$N_\mathrm{H}$ together with a softening of the underlying continuum.
If true, this overall phase dependence of $N_\mathrm{H}$ would imply
the presence of at most moderately ionized material that is located
close to the neutron star and coupled to its magnetic field. In our
model, we use a neutral absorption component, which is a common
approach in accreting X-ray pulsars. This is valid because most of the
absorption happens by K-shell electrons, so a moderately ionized
medium can still be successfully modelled with a neutral absorber. We
also applied the analytic \texttt{warmabs} model as part of the
\texttt{XSTAR}\footnote{\url{https://heasarc.gsfc.nasa.gov/docs/software/xstar/xstar.html}}
software package and found that an ionized absorber does not improve
our fit significantly in terms of $\chi^2$ and the resulting
ionization fractions are very low in both observations
($\log\xi\sim-0.05$).

While the general dependence of the spectral properties on pulse phase
is in line with results from the 2000/2001 outburst
\citep{Galloway2004,Naik2006}, our results differ in that no
$N_\mathrm{H}$ variability was seen in the older data. It is well
known that $N_\mathrm{H}$ in accreting neutron stars strongly depends
on the assumed underlying continuum model \citep[see, e.g., the
modeling of the $N_\mathrm{H}$ variability of
\object{4U\,1538$-$522};][]{Hemphill2014}. A possible explanation for
this discrepancy could therefore be that the pulse phase variability
of the earlier outburst was described by the variation of the optical
depth and temperature of a thermal Comptonization model
\citep[\texttt{compTT}][]{hua:95a} to which a soft black body was
added. This approach indeed resulted in much lower $N_\mathrm{H}$
values ($N_\mathrm{H}\sim5\times 10^{21}\,\text{cm}^{-2}$, i.e., the
minimum $N_\mathrm{H}$ detectable with the \rxte-PCA, but well within
the range accessible by \textsl{BeppoSAX};
\citealt{Galloway2004,Naik2006}).  Our attempts to model the 2013 data
with a Comptonization continuum resulted in barely acceptable fits
($\chi^2_\mathrm{red}(\mathrm{d.o.f}) = 1.43(715)$ for Obs.~I and
$\chi^2_\mathrm{red}(\mathrm{d.o.f}) = 1.56(561)$ for Obs.~II), and
even though $N_\mathrm{H}$ turned out to be comparable with the
2000/2001 values, the fits still showed a similar pulse phase
dependency of $N_\mathrm{H}$ to that found with the empirical models
employed in Sect.~\ref{sec:phase_resolved}.

A next step in describing the data would be to replace empirical
models such as our approach or single temperature Comptonization
models with a fully self-consistent description of the physical
processes in the accretion column \citep[e.g.,][and references
therein]{Becker2007}. The publicly available \texttt{XSPEC} model
\texttt{compmag} \citep{Farinelli2012} provides a numerical solution
of the equation of radiative transfer given in \citet{Becker2007}
modified by a second order bulk Componization term and allowing for
other velocity profiles than considered in \citet{Becker2007}. We have
tried to fit the \texttt{compmag} model to the phase-averaged spectra
which turned out not to describe the data successfully. We note,
however, the \texttt{compmag} model uses a blackbody seed spectrum to
be convolved with the Green's function, whereas in the full
\citet{Becker2007} model, also bremsstrahlung and cyclotron emission
contribute to the seed spectrum, which play a dominant role at high
luminosities. The \texttt{compmag} model is therefore rather
applicable to low luminosity observations and does not fit the bright
observations of \ks very well. Members of our team are currently in
the process of preparing an implementation of the full model of
\citet{Becker2007}. An independent implementation of the
\citet{Becker2007} model, together with a thermal Comptonization
component has been applied successfully to the spectrum of
\object{4U\,0115+634} by \citet{Ferrigno2009}.

For a full picture of the accretion mechanism, combined spectral and
timing analysis is required. The self-consistent modeling of both the
spectra and the pulse profiles will finally allow us to disentangle
artificial parameter correlations and track physical properties over
pulse phase. This is, however, a very challenging undertaking due to
large number of free parameters and because the required computing
time is very high in both the spectral and geometrical models. Huge
efforts in developing and improving these models have been made for
several years now and we are still at a point, where every successful
application to observational data provides a valuable example. \ks
seems to be a promising candidate for more dedicated studies and we
hope that additional constraints on the source geometry will
eventually help to improve our understanding of the spectral and
angular redistribution of radiation inside the accretion column.

\begin{acknowledgement}
  We thank the anonymous referee for very constructive comments that
  helped us to improve the quality of the paper. We acknowledge
  funding by the Bundes\-mi\-ni\-ste\-rium f\"ur Wirtschaft und
  Technologie under Deutsches Zentrum f\"ur Luft- und Raumfahrt grants
  50OR1113 and 50OR1207. V.G.\ acknowledges support provided by NASA
  through the Smithsonian Astrophysical Observatory (SAO) contract
  SV3-73016 to MIT for support of the Chandra X-Ray Center (CXC) and
  Science Instruments; CXC is operated by SAO for and on behalf of
  NASA under contract NAS8-03060. We thank John E.\ Davis for the
  development of the \texttt{SLXfig} module, which was used to create
  all figures presented in this paper. This research has made use of
  ISIS functions (\texttt{isisscripts}) provided by ECAP/Remeis
  observatory and
  MIT\footnote{\url{http://www.sternwarte.uni-erlangen.de/isis/}}.
\end{acknowledgement}

\bibliographystyle{aa}
\bibliography{references}

\end{document}